# Hydrogen-Bond Donor-Acceptor Imbalance in Low-Frequency Terahertz Water Spectra

Lilian Najm Alsayed[1], Florian Pabst[2], Giuseppe Cassone[3], Ali Hassanali[2], Fabio Novelli[1,*]

[1]School of Physics and Astronomy, Faculty of Engineering and Physical Sciences, University of Southampton, UK

[2]CMSP, The Abdus Salam Centre for Theoretical Physics, 34151 Trieste, Italy

[3]Institute for Chemical-Physical Processes, National Research Council of Italy, viale Ferdinando Stagno d'Alcontres 37, 98158 Messina, Italy

*F.Novelli@soton.ac.uk

*The low-frequency dielectric response of liquid water is commonly described by a dominant Debye relaxation together with additional faster contributions whose microscopic origin remains debated. Here we show that the dielectric function of water between 0.14 and 1.21 THz can be represented by a collective Debye relaxation plus a Drude-Smith term constrained to the zero-dc-conductivity limit. The Drude-Smith spectral weight increases upon heating pure liquid water from 20 °C to 50 °C and decreases upon isotopic substitution ($D_2O$ at 20 °C vs. $H_2O$ at 20 °C). Among several microscopic measures of electronic polarization and its fluctuations extracted from molecular-dynamics simulations, including path-integral simulations with nuclear quantum effects, the population of water molecules with unequal numbers of donated and accepted hydrogen bonds best tracks the trends of the Drude-Smith spectral weight, changing with the same sign and comparable magnitude. Because such donor-acceptor imbalance is the local condition for net intermolecular charge transfer, it points to fluctuating molecular charges, a degree of freedom that fixed-charge representations lack by construction. Consistently, ab-initio-based dielectric spectra show that the ~0.1–1 THz response cannot be solely described within a fixed-charge representation. These results support a plausible interpretation of parts of the excess low-frequency THz response of water as electronic-polarization fluctuations associated with transient donor-acceptor imbalance within the aqueous hydrogen-bond network.*

## INTRODUCTION

One of the key properties of the hydrogen bond is the transfer of a fraction of the electronic charge between donor and acceptor molecules, leading to partial covalent character [1]. This charge transfer term, which can contribute to ~20-40% of the bond energy in a water dimer [2,3], arises due to a small charge delocalization [4–6] involving the acceptor on one water molecule (oxygen lone-pair orbitals) and the nearby donor molecule (antibonding $\sigma^*_{OH}$ orbital) [7]. It has been suggested that such charge-transfer processes may affect the intensity [8–10] and peak positions [11,12] of the vibrational spectra of liquid water, contribute ~5-6% to the electronic polarizability of water in the bulk [13], and assist the stabilization of aqueous hydrophobic interfaces [14,15].

In the simplest, qualitative terms, if one water molecule in the liquid phase forms four hydrogen bonds, two donated and two accepted in a tetrahedral-like arrangement, the charge redistribution associated with the two accepted bonds is expected to compensate the charge redistribution associated with the two donated bonds. However, a substantial fraction of water molecules in the liquid phase are transiently under- or over-coordinated at any instant of time, and several of these non-tetrahedral environments involve an imbalance between the number of donated and accepted hydrogen bonds [16–18]. For molecules with such an instantaneous imbalance, both the extent of charge-transfer and its contribution to the electronic polarizability are expected to vary. This local asymmetry, in turn, could contribute to dielectric fluctuations with characteristic spectroscopic signatures.

In simple conductors, the Drude model describes the spectroscopic response in terms of free carriers characterized by a plasma frequency and a scattering time. Whenever the field-induced current is spatially confined, for example in nanostructured or disordered media [19–21], the Drude description becomes inadequate, and the semi-empirical Drude-Smith formalism can be used [22–28]. In the original formulation [29], Smith introduced "backscattering" to account for the possibility that the carrier velocity reverses direction before decaying. For example, in a nanomaterial, the carriers could bounce off the boundary and reverse direction of motion [19–21]. Cocker *et al.* [30] later performed Monte Carlo simulations and suggested that the Drude-Smith response stems from diffusive restoring currents, implying hindered, more localized transport. Moreover, while Kužel and Němec [31] stressed that the Drude-Smith is an effective model that requires independent microscopic validation, Chen and Marcus [32] derived it under the assumption of non-Markovian transport, that is, whenever the transport cannot be written as a function of a single temporal variable but depends on the history of the charge interactions.

Liquid water displays a broad spectroscopic feature centered at ~20 GHz and extending into the terahertz (THz) frequency range, which often fits a Debye function with a reorientation time constant of 8-10 ps [33–37]. In the original formulation, the Debye function describes the time it takes for pre-aligned and non-interacting dipoles to randomize after the driving electric field is turned off [38]. More broadly, the Debye function describes stochastic processes with an effectively instantaneous memory kernel, implying that the collective polarization can be described by Markovian dynamics [32]. Above about 0.2 THz, however, a single Debye term is insufficient to describe the dielectric function of water, and at least one additional contribution is required [39,40]. While it could be preferable to use a broad distribution rather than one additional discrete process [41–45], several fit procedures were proposed by adding either a log-normal function describing excess vibrational states akin to the boson peak in glass-forming materials [46,47], Lorentzian-shaped Raman responses owing to the fact that a water molecule lacks a center of inversion [48], overdamped oscillators [49], or Debye function(s) with faster reorientation time constants of 0.1-1 ps.

The latter description of the THz response of water with two Debye functions is widely used [50–53]. However, while the main Debye relaxation can be associated with collective dipolar reorientation of hydrogen-bonded water molecules [38,54,55], the microscopic description of the faster Debye term(s) remains unclear. These faster Debye-type processes have been associated with water molecules with weaker hydrogen bonds [33,50,51,56]. If this would be the case we expect, to a first approximation, clear temperature and isotopic trends: deuterated molecules should be slower due to inertial effects and a more rigid hydrogen-bond network, which stems from mitigated nuclear quantum effects; and there should be more weakly hydrogen-bonded molecules at higher temperatures. These expectations disagree with the double-Debye fit results. For example, Figure 2 of the seminal work by Rønne *et al*. [34] indicated that the second Debye reorientation time-constant can be *faster* in $D_2O$ than in $H_2O$, rather than slower. More recently, Figure 3 in Yada *et al*. [36], Figure 2 in Kutus *et al*. [37], and Figure 5 in Taschin *et al.* [45] reported negligible temperature and isotope effects.

Here we show that a fit including the main Debye term with 8-10 ps time-constant and one Drude-Smith component describes the dielectric response of liquid water between about 0.14 and 1.21 THz. We recorded spectra of $H_2O$ at 20 °C, $H_2O$ at 50 °C, $D_2O$ at 20 °C, and found that the Drude-Smith effective spectral weight increases when the hydrogen-bond network is weakened by heating and decreases when it is strengthened by deuteration. This provides an empirical approach to fitting the low-frequency THz spectra of aqueous systems, in which the fit parameters correlate with average structural properties of the hydrogen-bond network.

To investigate the microscopic origin of the Drude-Smith spectral weight, we use dedicated molecular dynamics simulations based on neural network potentials (NNP) and quantify charge redistribution and dipole fluctuations in different hydrogen bond environments. We then compare their temperature and isotope trends with those of the fitted Drude-Smith spectral weight. The measure that best tracks the experimental trends is the population of molecules with unequal numbers of donated and accepted hydrogen bonds. This is the local condition under which intermolecular charge transfer does not cancel, and a molecule acquires a transient net charge. Since such charge fluctuations are absent by construction in a fixed-charge representation, we use *ab-initio*-based calculations on the NNP trajectories to isolate the contribution of electronic polarization to the dielectric function. While it is established that polarization effects must be included to describe the THz response of water above about 1 THz [8–10,57,58], our results indicate that electronic polarization is already relevant in the ~0.1–1 THz range, which overlaps with the experimental Drude-Smith component.

# METHODS

## Terahertz spectroscopy

Terahertz radiation was generated by tilted-front optical rectification of ~90-fs long titanium-sapphire amplified pulses in a lithium niobate prism. The transmission by the opaque samples was detected via electro-optical sampling in a 1 mm thick zinc telluride crystal, over a 13.9 ps temporal range, and in steps of 0.1 ps. The acquisition of one THz trace took about 20 seconds and was repeated 15 times. Uncertainties from these repetitions were estimated as 95% confidence intervals (CI). The samples were mounted in a static liquid cell with 0.5 mm thick diamond windows, a 0.5 mm Teflon spacer, and attached to a recirculating chiller stabilizing the temperature within ±0.05 °C. The THz spectrometer was purged with nitrogen to reduce the relative humidity to below 10%. Pure milli-Q liquid water at 20±0.05 °C was used as reference, see the black curve in Figure 1. As shown in Figure 1 inset, the corresponding magnitude spectrum spans between about 0.14 and 1.21 THz with a dynamic range of ~30 dB. We also measured pure milli-Q liquid water at 50±0.05 °C (red in Figure 1) and >99%-pure $D_2O$ at 20±0.05 °C (blue in Figure 1).

The optical properties of these samples were obtained by numerical inversion of Eq. (3) in Ref. [59], which reads

$$\frac{E_{sample}}{E_{wat}} = \frac{(n_{sample}+ik_{sample})(n_{wat}+ik_{wat}+2.37)^2}{(n_{wat}+ik_{wat})(n_{sample}+ik_{sample}+2.37)^2} e^{i\frac{\omega d}{c}(n_{sample}-n_{wat})} e^{-\frac{\omega d}{c}(k_{sample}-k_{wat})} \quad (1)$$

where $i$ is the imaginary unit, $E_{sample}$ is the complex and frequency-dependent Fourier-transform (FT) of the THz field transmitted by a sample ($H_2O$ at 50 °C or $D_2O$ at 20 °C), $E_{wat}$ is the FT of the reference (pure $H_2O$ at 20 °C), $n_{wat}$ and $k_{wat}$ are known [59], 2.37 is the real index of refraction of the diamond windows of the sample holder [59], and $n_{sample}$ ($k_{sample}$) is the sought index of refraction (extinction coefficient). Further details on intense THz time-domain spectroscopy can be found in Refs. [59–61].

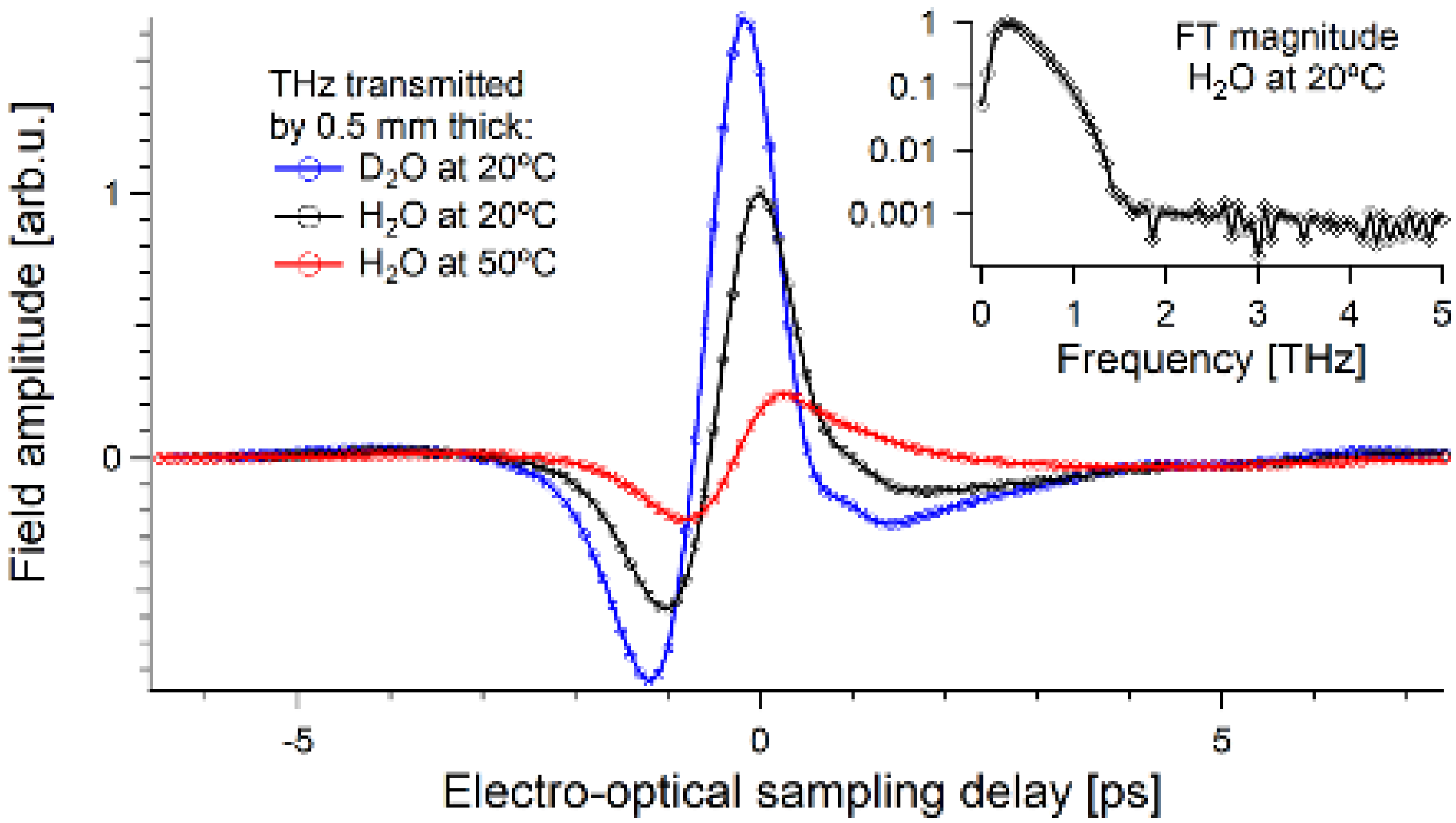

***Figure 1**. Intense time-domain terahertz spectroscopy of water with different hydrogen-bonding strength. The THz field transmitted by a 0.5 mm thick layer of pure water at 20 °C is shown with black circles and the corresponding spectrum is reported in the inset with black diamonds. As shown by the blue THz trace, both absorption and refraction decrease by isotope substitution ($D_2O$). On the contrary, when $H_2O$ is heated, it becomes more absorptive and refractive: see the red trace that is obtained at the temperature of 50 °C. The 95% CI calculated from 15 measurements are plotted together with each THz trace, but they are not visible because they are smaller than the line thickness.*

**Dielectric models**

The real and imaginary parts of the index of refraction were calculated from the known relationships $2n^2 = \varepsilon_1 + \sqrt{\varepsilon_1^2 + \varepsilon_2^2}$ and $2k^2 = -\varepsilon_1 + \sqrt{\varepsilon_1^2 + \varepsilon_2^2}$, respectively. For the double-Debye representation, we used

$$\varepsilon_1(\omega) = \varepsilon_\infty + \frac{\Delta\varepsilon_1}{1+\omega^2\tau_1^2} + \frac{\Delta\varepsilon_2}{1+\omega^2\tau_2^2} \tag{2}$$

$$\varepsilon_2(\omega) = \frac{\Delta\varepsilon_1\omega\tau_1}{1+\omega^2\tau_1^2} + \frac{\Delta\varepsilon_2\omega\tau_2}{1+\omega^2\tau_2^2} \tag{3}$$

where $\omega = 2\pi\nu$ is the angular frequency, $\varepsilon_\infty$ dielectric constant at higher frequencies, $\Delta\varepsilon_1 = \varepsilon_S - \varepsilon_d$ and $\Delta\varepsilon_2 = \varepsilon_d - \varepsilon_\infty$ the weights or amplitudes of the Debye-type reorientations with time-constants equal to $\tau_1$ and $\tau_2$, respectively, and $\varepsilon_S$ static dielectric constant fixed to literature values [62–64].

For the Debye plus Drude-Smith fit, the following functions were used:

$$\varepsilon_1(\omega) = \varepsilon_\infty + \frac{\Delta\varepsilon_1'}{1+\omega^2\tau_1^2} - \frac{\omega_P^2\tau^2(1+\omega^2\tau^2+2c)}{(1+\omega^2\tau^2)^2} \tag{4}$$

$$\varepsilon_2(\omega) = \frac{\Delta\varepsilon_1'\omega\tau_1}{1+\omega^2\tau_1^2} + \frac{\omega_P^2\tau[1+\omega^2\tau^2+c(1-\omega^2\tau^2)]}{\omega(1+\omega^2\tau^2)^2} \tag{5}$$

including the effective localization parameter $c$, effective plasma frequency $\omega_P$, and effective scattering time $\tau$. The strength of the Debye term is determined by $\Delta\varepsilon_1'$ and constrained by the limiting values of the dielectric function for static and high-frequency fields. By rearranging the identity $\varepsilon_S = \varepsilon_1(0) = \varepsilon_\infty + \Delta\varepsilon_1' - \omega_P^2\tau^2(1+2c)$, we obtain $\Delta\varepsilon_1' = \varepsilon_S - \varepsilon_\infty - \omega_P^2\tau^2$ for $c = -1$.

While plasma frequency and scattering time are present in both Drude and Drude-Smith models, the $c$ parameter determines the effective physics underlying the transport regime. For $c = 0$ there is no backscattering or charge localization, and the Drude-Smith is identical to the Drude model with long-range carrier transport and non-zero conductivity at zero frequency. For $c = -1$ the localization is complete, the static conductivity is reduced to zero, and no long-range transport is possible. As pure liquid water does not support long-range dc transport by free electronic carriers, here we explore whether its low-frequency response can be described in part by a Drude-Smith term with $c$ fixed to $-1$. This also ensured that the double-Debye and Debye plus Drude-Smith fits have the same number of free parameters: $\varepsilon_d, \varepsilon_\infty, \tau_1, \tau_2$ for the double-Debye; $\varepsilon_\infty, \tau_1, \omega_P, \tau$ for the Debye plus Drude-Smith.

If $c$ is not fixed to $-1$, it converges to best fit values between about $-1.1$ and $-1.2$, which imply negative dc conductivity and are therefore unphysical. However, even in this case, we found that the overall trend of the plasma frequency as a function of temperature and isotopic substitution is maintained, thereby leaving the main conclusions unchanged. Please note that, for $c = -1$, the Drude-Smith equations are identical to the ones describing a critically damped harmonic Lorentz oscillator [65] with resonance frequency equal to $1/\tau$ and damping rate of $2/\tau$.

To estimate the fit quality, we calculated the sum of the absolute differences between the fit results and the original data. We divided this sum by the average response between 0.14 and 1.21 THz. In other words, we calculated $\bar{n}_{res} = \sum_\nu |n_{res}(\nu)| / \sum_\nu n(\nu)$ for the real part of the refractive index, where each residue $n_{res}(\nu)$ is equal to the difference between the data and the fit at frequency $\nu$. Likewise, for the extinction coefficient we used $\bar{k}_{res} = \sum_\nu |k_{res}(\nu)| / \sum_\nu k(\nu)$. The results are summarized in the last row of Table I and Table II.

**Molecular dynamics simulations**

Our strategy to compare with the experimental measurements leverages two complementary water models based on NNP. Firstly, to understand the role of isotope effects, one needs to explicitly include nuclear quantum effects in the simulations, which is computationally challenging. We thus deployed a computationally efficient deep-NNP including nuclear quantum effects with path-integral molecular dynamics (PI-DPMD). For the temperature dependence comparing two state points, one at a higher and one at a lower temperature, we deployed another NNP trained on data coming from the MB-pol water model. Details of both models are summarized next.

Isotope effect. Neural-network-based molecular dynamics simulations were carried out with the LAMMPS software (v. 29Aug2024) [66]. A computationally efficient compressed model was used that yields better performance. In particular, the employed reactive deep potential was trained on approximate hybrid meta-GGA M06-2X [67] data, which was reported and validated in Refs. [68,69]. The isotope comparison was evaluated by performing simulations within the path integral formalism recently introduced by Car and co-workers [70] both for $H_2O$ and $D_2O$ in the canonical NVT ensemble at 298 K. These models consist of 256 water molecules in cubic boxes with edges of 19.641 Å. The PI-DPMD simulations were performed at a density of 1.0 g/cm$^3$ for $H_2O$ and 1.11 g/cm$^3$ for $D_2O$. They were run for 500 ps with a timestep of 0.25 fs.

Temperature effect. Temperature-dependent simulations with the MB-pol deep neural network [71] were performed using LAMMPS [66], interfaced with the DeepMD-kit [72], on bigger systems of 512 water molecules. First, NPT simulations were performed to equilibrate the system with a time step of 0.5 fs. The density obtained at the end of this simulation was then used for a subsequent NVT equilibration run, again with a 0.5 fs time step. From this run, eight independent configurations were extracted and used as starting points for eight production runs in the NVE ensemble, with a time step of 0.25 fs and a duration of 5 ns at 17 °C and 2 ns at 57 °C.

Dielectric function. Along the MB-pol trajectories described above, the imaginary part of the dielectric response of $H_2O$ at 57 °C was calculated from the total electric dipole moment $\mathbf{M}$ of the simulation box. The molecular dipoles were obtained either using an independently trained neural-network dipole model reproducing DFT-based Wannier-center dipoles trained with the SCAN exchange-correlation functional [73,74], or by placing fixed charges of +0.5e and -1e on the hydrogen or oxygen atom, respectively, via the equation

$$\varepsilon_2(\omega) = \frac{\omega}{3V\varepsilon_0 k_B T} \int dt\, e^{-i\omega t} \langle \mathbf{M}(t) \cdot \mathbf{M}(0) \rangle \qquad (6)$$

where $V$ is the box volume, $\mathbf{M}$ is the total dipole moment of the system (i.e., the sum over all molecular dipole moments), $T$ is the simulation temperature, and $\varepsilon_0$ and $k_B$ are the vacuum permittivity and Boltzmann constant, respectively. The autocorrelation functions of the dipole moment time series from the eight production runs are averaged to reduce the noise of the spectra. To allow for a quantitative comparison between the spectral shapes obtained with the two approaches, a small scaling factor (1.036) was applied to the fixed-charge spectrum to overlap the spectra at the main Debye peak, compensating for the arbitrary absolute dipole magnitude imposed by the fixed partial charges.

Hydrogen-bond analysis. Hydrogen bonds were identified using the geometric criteria developed by Luzar and Chandler where an O···O distance ≤3.5 Å and an H–O···O angle below 30° is considered a hydrogen bond [75]. We classified each instantaneous molecular environment by the number of donated and accepted hydrogen bonds since it has been previously shown that the asymmetry in the hydrogen-bond local topology leads to charge transfer [17,18]. All the configurations of the water molecules as a function of the number of donated and accepted hydrogen bonds are reported in Supplementary Table 1 and Supplementary Table 2 for all the simulations. Because the isotope and temperature trends associated with different hydrogen-bond populations were obtained using different simulation protocols and are clearly model-dependent, we interpreted only relative changes within each simulation protocol.

Wannier Analysis. Besides the structural parameters associated with the hydrogen-bond imbalance, we also examined several other measures that probe electronic polarization. Specifically, we used the deep neural network trained to reproduce the Wannier dipole [73] to extract the electronic dipole distributions associated with the water molecules. Besides the dipole, we also examined how the distribution of the spread of the maximally localized Wannier functions [76] changes as a function of isotope and temperature. For this, we extracted 200 frames from the different conditions and calculated the Wannier spreads using the program CP2K and the SCAN functional.

## RESULTS

The dielectric function of liquid water at THz frequencies and 20 °C was tabulated by collating multiple previous reports, see Table 1 in Ref. [59]. The corresponding refractive index ($n_{wat}$) and extinction coefficient ($k_{wat}$) are shown as blue and red circles in Figure 2, respectively.

We initially performed a fit of the dielectric function of pure water at 20 °C with the standard double-Debye described by Eqs. (2) and (3), and compared it with the single Debye plus Drude-Smith fit, see Eqs. (4) and (5). The results are reported in Figure 2a and Figure 2b, respectively. The black lines are the fit results of the double-Debye fit; the gray ones are Debye plus Drude-Smith results. The two components of the extinction coefficient are sketched at the bottom with the usual, main Debye-type reorientation (dark green in Figure 2) dominating the response at lower THz frequencies. The remaining “bump at higher frequencies” can be represented by either a second Debye term (light green in Figure 2a) or the Drude-Smith component (purple in Figure 2b). As shown on top of Figure 2 and in the last row in Table I and Table II, the magnitude of the residuals of the Debye plus Drude-Smith fit are comparable to the ones obtained with the well-established double-Debye approach.

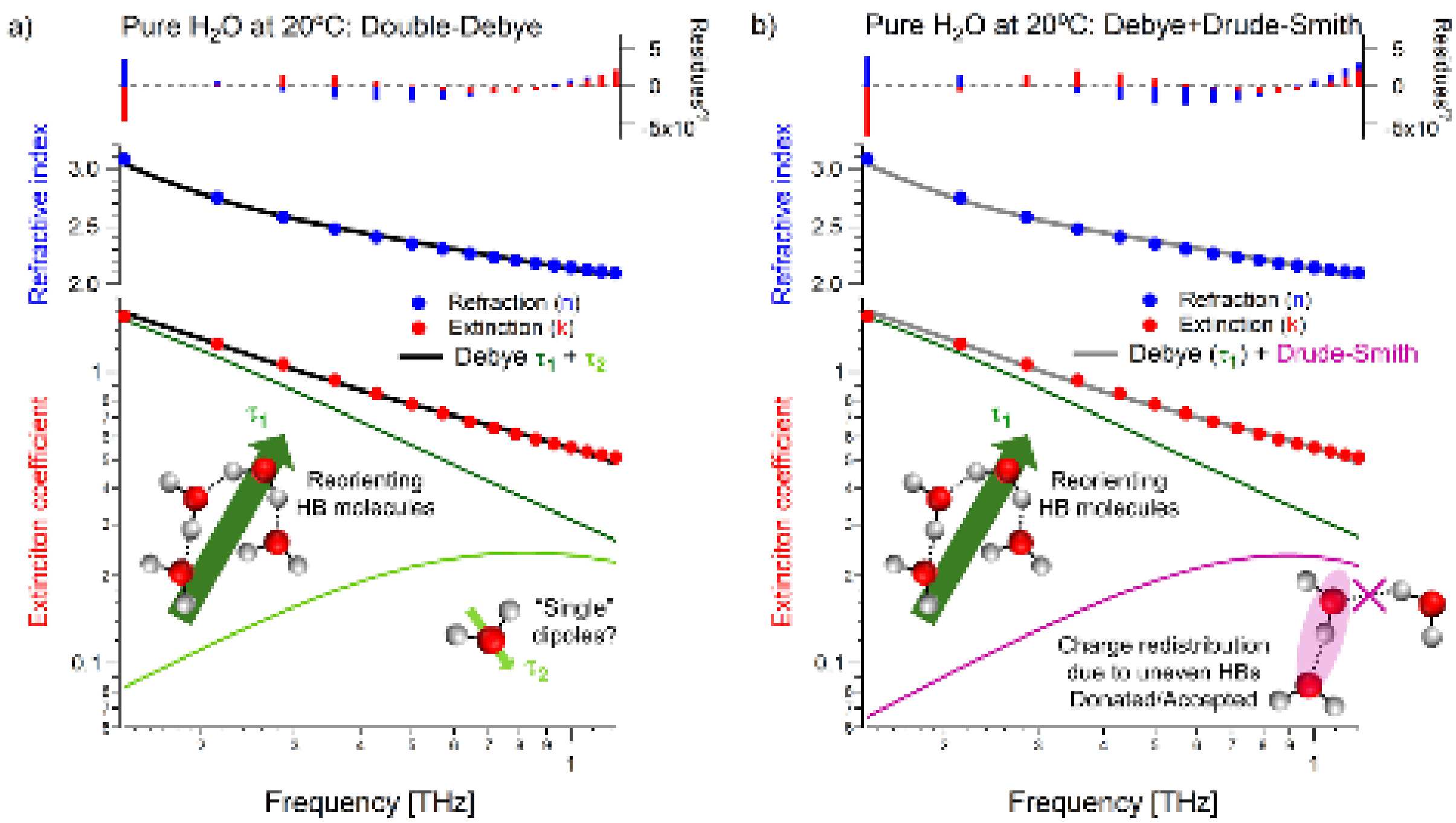

***Figure 2***. *Fitting the low-frequency THz dielectric function of liquid water at 20 °C estimated from previous reports [59]. The fit residues are shown on top, the real part of the refractive index of water is reported in blue, and the extinction coefficient in red. a) The usual double-Debye fit (black lines) is consistent with two reorientation mechanisms with time-constants of about 8 ps and 200 fs, respectively. While the slower and dominant Debye process is consistent with dipole reorientations of groups of hydrogen-bonded water molecules, the faster and weaker process has been ascribed to water molecules characterized by weaker or distorted hydrogen bonding. b) A single Debye plus Drude-Smith fit (gray lines) produces comparable results. While the slower and dominant Debye process remains, the higher-frequency response is empirically consistent with localized dielectric response associated with imbalanced donor/acceptor hydrogen-bond environments.*

Both the double Debye and the Debye plus Drude-Smith fits can therefore reproduce the experiments on pure water at 20 °C. Based on this one state point we cannot choose one fit or another in terms of which captures the experimental trends better. We thus turned to exploring how well these models extend to other thermodynamic conditions, i.e., we weakened the hydrogen-bond network by heating or strengthened it via deuteration at room temperature [11]. In Figure 3 we display the refractive (blue) and extinction (red) coefficients: triangles are used for $H_2O$ at 50 °C, squares for $D_2O$ at 20 °C. The Debye plus Drude-Smith fit is shown in gray in Figure 3 and the residues, which are reported at the top of Figure 3 with vertical bars, are larger than the measurement uncertainties (horizontal steps).

By comparing the fit results in Table I and Table II, we notice that the time-constants of the main Debye terms are similar across different fit procedures: $\tau_1$ is about 8.6 ps for $H_2O$ at 20 °C, 4.7 ps for $H_2O$ at 50 °C, and 10.9 ps for $D_2O$ at 20 °C. These results agree with previous reports detailing faster time constants when water is heated and slower ones upon deuteration [33–37]. However, the fit parameters describing the THz excess have different trends across the two models used. For the second Debye term, the parameters depend weakly on temperature and isotopic substitution, in agreement with several previous reports, including Refs. [34,36,37]. In Table I, the reorientation time-constant is close to 200 fs while its amplitude, $\Delta\varepsilon_2 = \varepsilon_d - \varepsilon_\infty$, is close to 2. The mean fit value for $\Delta\varepsilon_2$ varies erratically: it decreases when water is heated and also upon isotopic substitution. Therefore, it cannot simply correlate with the hydrogen-bond strength or the population of hydrogen-bond defects. The Drude-Smith term displays a scattering time close to 100 fs in all cases investigated and, importantly, a clear trend for the plasma frequency: it increases upon heating and decreases upon isotopic substitution. This fit parameter therefore is qualitatively consistent with the hydrogen-bond network strength that, when comparing to $H_2O$ at 20 °C, becomes weaker in $H_2O$ at 50 °C, and stronger in $D_2O$ at 20 °C.

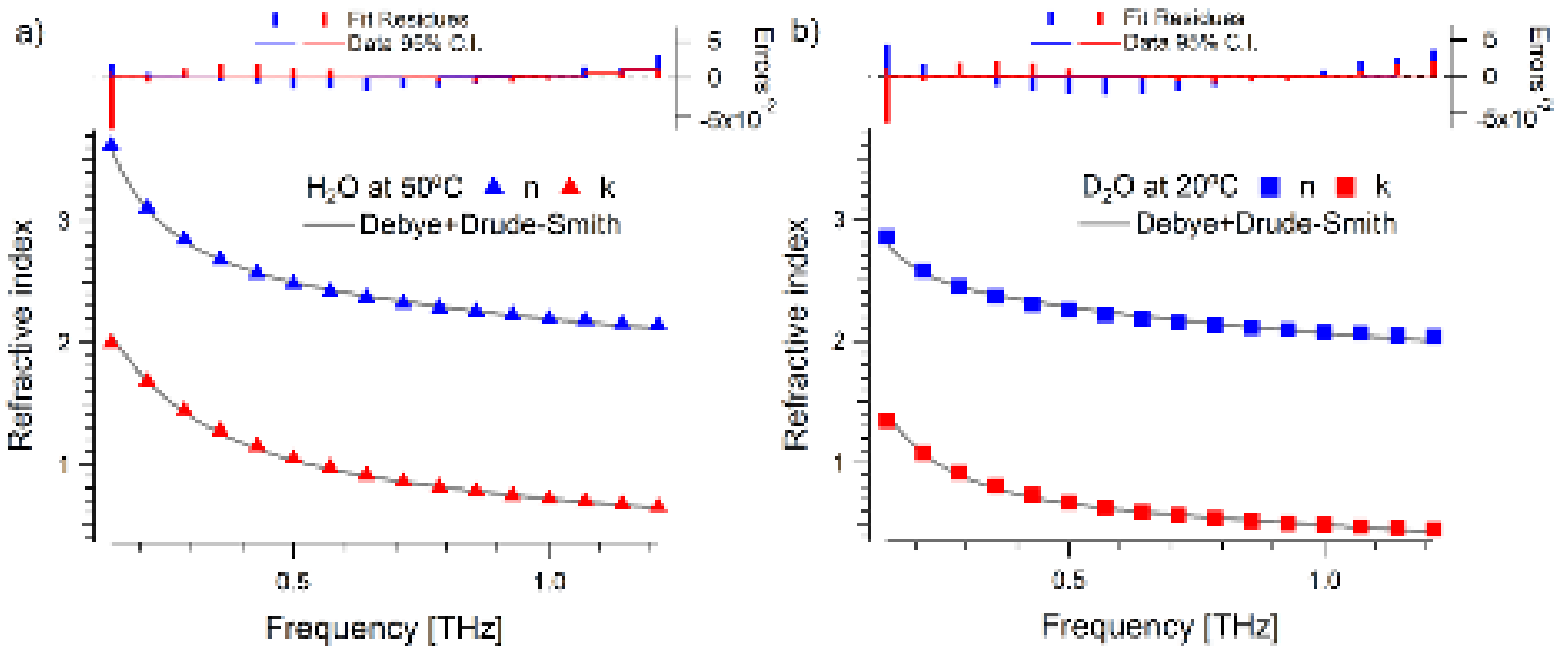

***Figure 3****. Real and imaginary parts of the complex index of refraction of pure $H_2O$ at 50 °C (a) and of heavy water at 20 °C (b). As expected, when compared to pure water at 20 °C, $D_2O$ displays a smaller complex refractive index, while heating is associated with a stronger field-matter interaction. The single Debye plus Drude-Smith fit is shown with gray lines. The fit residues are reported by vertical bars in the top panel. The 95% CI from the measurements are reported with horizontal segments and are smaller than the fit residues.*

***Table I****. Fit results with the standard double Debye fit. The static dielectric constants ($\varepsilon_S$) were fixed to literature values [62–64]. $\varepsilon_d$ is the first Debye term (sometimes labeled with $\varepsilon_1$), $\varepsilon_\infty$ the dielectric constant at higher frequencies, $\tau_1$ (sometimes labeled with $\tau_D$) and $\tau_2$ are the reorientation time-constants of the first and second Debye terms, respectively.*

| | **$H_2O$ at 20 °C** | **$H_2O$ at 50 °C** | **$D_2O$ at 20 °C** |
|---|---|---|---|
| $\boldsymbol{\varepsilon_S}$ | 80.2 | 69.9 | 79.8 |
| $\boldsymbol{\varepsilon_d}$ | 5.57 ± 0.07 | 5.37 ± 0.07 | 5.32 ± 0.07 |
| $\boldsymbol{\varepsilon_\infty}$ | 3.53 ± 0.06 | 3.43 ± 0.07 | 3.35 ± 0.06 |
| $\boldsymbol{\Delta\varepsilon_2 = \varepsilon_d - \varepsilon_\infty}$ | 2.04 ± 0.04 | 1.94 ± 0.05 | 1.97 ± 0.04 |
| $\boldsymbol{\tau_1}$ **[ps]** | 8.72 ± 0.09 | 4.71 ± 0.04 | 11.02 ± 0.13 |
| $\boldsymbol{\tau_2}$ **[fs]** | 219 ± 13 | 189 ± 13 | 222 ± 13 |
| $\boldsymbol{\bar{n}_{res}}$ **[%] ;** $\boldsymbol{\bar{k}_{res}}$ **[%]** | 0.6 ; 1.4 | 0.4 ; 1.0 | 0.6 ; 1.6 |

***Table II**. Fit results with the single Debye plus Drude-Smith. The static dielectric constants were fixed to literature values [62–64]. The $c$ parameter was fixed to a value equal to $-1$ to account for full localization [20,30]. $\omega_P$ is the effective plasma frequency and $\tau$ effective scattering time.*

| | $H_2O$ at 20 °C | $H_2O$ at 50 °C | $D_2O$ at 20 °C |
|---|---|---|---|
| $\varepsilon_S$ | 80.2 | 69.9 | 79.8 |
| $\varepsilon_\infty$ | 3.87 ± 0.03 | 3.77 ± 0.04 | 3.70 ± 0.01 |
| $\tau_1$ [ps] | 8.50 ± 0.05 | 4.66 ± 0.02 | 10.77 ± 0.13 |
| c | -1 | -1 | -1 |
| $\omega_P$ [rad/ps] | 11.16 ± 0.31 | 12.13 ± 0.44 | 10.60 ± 0.11 |
| $\tau$ [fs] | 110 ± 3 | 100 ± 4 | 114 ± 1 |
| $\bar{n}_{res}$ [%] ; $\bar{k}_{res}$ [%] | 0.7 ; 1.6 | 0.5 ; 1.0 | 0.8 ; 1.8 |

DISCUSSION

In a conventional Drude model, $\omega_P^2$ is proportional to the density of free carriers. In the present context, this microscopic interpretation is not obvious, because pure liquid water is a large band-gap insulator and does not support long-range electronic transport (i.e., it is an electronic insulator). However, by looking at the values of the plasma frequency in Table II, we found a qualitative scaling with the hydrogen-bond strength. As several previous theoretical studies have shown that asymmetries in the hydrogen-bonding environments created by thermal fluctuations can lead to imbalances in electronic polarization and subsequently a net charge-transfer [3,15,77–79] , we speculate that $\omega_P^2$ might correlate with the fraction of water molecules in an asymmetric environment, wherein the local charge redistribution is imbalanced due to an uneven number of donated and accepted hydrogen bonds. In particular, from the values listed in Table II, the effective spectral weight $\omega_P^2$ increases by $\frac{\omega_P^2(50°C)-\omega_P^2(20°C)}{\omega_P^2(20°C)} \approx +(18 \pm 8)\%$ in water at 50 °C with respect to water at 20 °C, and decreases by $\frac{\omega_P^2(D_2O)-\omega_P^2(H_2O)}{\omega_P^2(H_2O)} \approx -(10 \pm 4)\%$ when comparing $D_2O$ at 20 °C with $H_2O$ at 20 °C.

To assess the validity of assuming that the Drude-Smith spectral weight might be related to donor/acceptor-imbalanced environments with asymmetric charge redistributions, we searched for an empirical correlation with the simulated population of water molecules having unequal numbers of donated and accepted hydrogen bonds. According to the path integral deep potential molecular dynamics simulation results that include nuclear quantum effects, the fraction of water molecules with an uneven number of donated and accepted bonds decreases from 19.18% in $H_2O$ at 298 K to 16.57% in $D_2O$ at 298 K (see the "Sum D ≠ A" row at the bottom of Supplementary Table 1). The corresponding variation is $16.57/19.18 - 1 \sim -13.6\%$. This correlates with the experimental Drude-Smith estimate, $-(10 \pm 4)\%$. According to simulation results of pure water at different temperatures, we estimated that heating $H_2O$ from 17 °C to 57 °C increases the fraction of water molecules with an uneven number of accepted and donated bonds from 37.68% to 45.92%. The corresponding variation is $45.92/37.68 - 1 \sim +21.9\%$, which correlates with the experimental Drude-Smith estimate of $+(18 \pm 8)\%$. These correlations between simulated microscopic structural properties and experimental dielectric fit parameters are noteworthy because previous attempts with an additional Debye component have not provided consistent trends across temperature and isotope changes [34,36,37].

The correlation with the hydrogen-bond populations described above provides only indirect support for the proposed mechanism. Charge transfer is not itself a directly measurable experimental observable and, strictly speaking, is quantum-mechanically ill-defined; rather, it offers a microscopic, useful interpretation of enhanced spatial electronic polarization, whose fluctuations contribute to the dielectric response. To characterize this polarization, we used maximally localized Wannier functions to calculate molecular dipole moments and orbital spreads, resolving the latter into lone-pair and bonding-pair contributions. Their mean values show no appreciable dependence on temperature or isotopic substitution, whereas their fluctuations exhibit more pronounced changes (Table III). The variance of the molecular dipole increases upon heating and decreases upon deuteration, following the same directional trends as the donor-acceptor-imbalanced hydrogen-bond population and the fitted Drude-Smith spectral weight. Its relative changes are of a comparable order of magnitude to those observed experimentally but agree less closely than the changes in the population of asymmetric/imbalanced hydrogen bonds. By contrast, the variances of the Wannier spreads remain unchanged at higher temperature but decrease upon isotopic substitution. Molecular dipole fluctuations therefore qualitatively reproduce the trends inferred from the empirical fits, although the donor/acceptor imbalance remains the descriptor that best captures both the temperature and isotope effects.

***Table III.*** *Relative changes in the experimental Drude-Smith spectral weight and simulation-derived hydrogen-bond and electronic-polarization descriptors. All values are percentage variations.*

| | **Drude-Smith $\omega_P^2$ (exp.)** | **H-Bond Imbalance** | **Wannier Dipole Variance** | **Wannier-Bonding Spread Variance** | **Wannier-Lone Pair Spread Variance** |
|---|---|---|---|---|---|
| **$H_2O$, 17 → 57 °C** | +18 ± 8 | +21.9 | +9.9 | −0.6 | −0.3 |
| **$H_2O$ → $D_2O$, 298 K** | −10 ± 4 | −13.6 | −20.9 | −24.7 | −19.1 |

To further clarify the microscopic nature of the THz response of liquid water, we performed additional *ab-initio*-based simulations to calculate the imaginary part of the dielectric function of $H_2O$ at 57 °C. The results are shown in Figure 4. The simulations were performed either using a fixed partial-charge dipole approximation ($\varepsilon_{2,pc}$), which excludes explicit electronic polarization and charge redistribution from the dipole definition, or using an additional neural network trained to reproduce the molecular dipole moment based on a Wannier-center approach ($\varepsilon_{2,wan}$), which includes polarization effects. These results are shown by the red and blue curves in Figure 4a for the partial-charge and Wannier-center approaches, respectively.

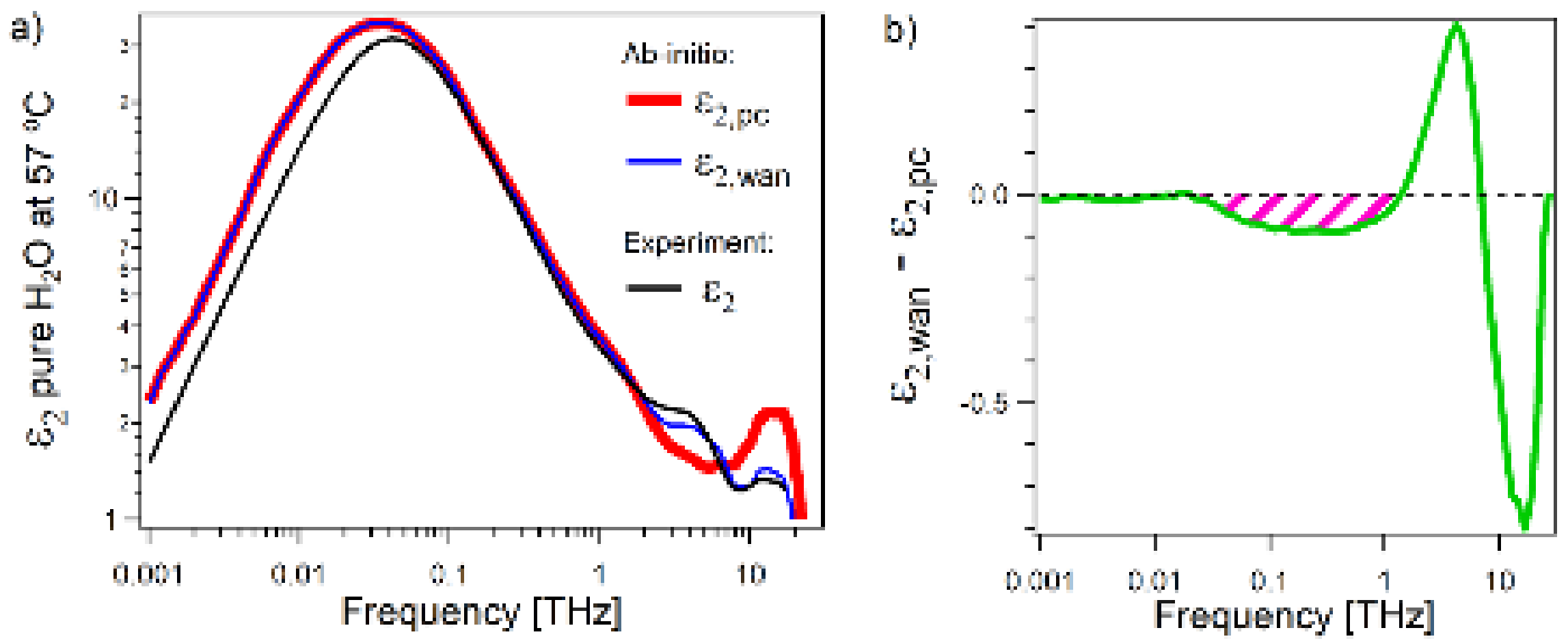

***Figure 4****. Imaginary part of the dielectric function of liquid water over a broadband terahertz spectral range. a) Comparison between established literature experimental results (black, from Refs.* [33,80]*) and ab-initio-based calculations including polarization effects ($\varepsilon_{2,wan}$ in blue) or a fixed partial-charge dipole approximation, which excludes explicit electronic-polarization and charge-redistribution from the dipole definition ($\varepsilon_{2,pc}$ in red). b) The green trace shows the difference between the Wannier-center and fixed-charge spectra, highlighting the part of the response not captured by the fixed-charge dipole approximation. This difference deviates from zero above about 0.03 THz, with a noticeable signal between about 0.1 and 1 THz that is highlighted by the purple shaded area. Large deviations are found in the intermolecular stretch and libration regions around ~5 and 17 THz, respectively.*

The green curve in Figure 4b corresponds to the difference between the Wannier-center and fixed-charge spectra, which were both calculated from the same MB-pol-based nuclear trajectories using different molecular-dipole representations. This difference highlights the part of the dielectric response that is not captured by the fixed-charge dipole approximation alone, including explicit electronic polarization and cross-correlation contributions. The fixed-charge model is inappropriate above about 1 THz, where the broad translational and librational features are evident. This is expected, because polarization effects are known to be relevant for these intermolecular hydrogen-bond-network modes [8–10,57,58]. The most interesting feature revealed from this analysis is that the fixed-charge estimate deviates already at lower frequencies, between about 0.1 and 1 THz, as highlighted by the purple-shaded area in Figure 4b. These simulations therefore support the notion that explicit electronic polarization contributes to the low-frequency THz response of liquid water. We note that this ~0.1-1 THz range also overlaps with the same spectral region that can be described by the empirical Drude-Smith fit.

## CONCLUSIONS

We investigated the low-frequency dielectric response of liquid water in the terahertz range by comparing the standard double-Debye approach with a model combining one Debye relaxation and one fully localized Drude-Smith contribution. Both approaches reproduce the main experimental features of the dielectric response of water between about 0.14 and 1.21 THz with comparable fit residuals. However, the Drude-Smith formalism allows for a plausible and complementary physical interpretation of the dielectric response of water.

The main Debye relaxation times are slower in heavy water and faster upon heating, which is qualitatively consistent with recent *ab-initio*-based results that associate the 8-10 ps Debye-type reorientation response to cross-dipolar correlations involving the orientations of multiple water molecules [54,55]. Importantly, the squared plasma frequency of the Drude-Smith fit correlates with the simulated population of water molecules having unequal numbers of donated and accepted hydrogen bonds. In fact, upon either deuteration or heating, the Drude-Smith spectral-weight changes have the same sign and comparable magnitude as the simulated changes in the donor/acceptor-imbalanced population. We show further that the fluctuations of the Wannier dipole, which more broadly capture electronic polarization, also qualitatively reproduce the same trends captured by the Drude-Smith weights as a function of temperature and isotope effects, although less closely than the local hydrogen-bond imbalance.

These results suggest that transient intermolecular charge redistribution associated with hydrogen-bond-network disorder could contribute to the low-frequency terahertz response of water. This interpretation is consistent in part with previous works linking the dielectric response of water to the migration of bifurcated hydrogen-bond defects [81], charge fluctuations [82,83], or hydrogen-bond lifetimes [84]. The advantages of the novel Drude-Smith description proposed here include an intuitive microscopic interpretation linked to charge localization, which is consistent with molecular dynamics simulations quantifying the fraction of water molecules characterized by an imbalance in donating versus accepting hydrogen bonds, and, importantly, a simple analytical form as shown in Eqs. (4) and (5). This practical approach could complement standard double-Debye procedures, and, in some cases, it might provide deeper microscopic insights.

Extending this Drude-Smith approach to electrolyte or protein buffer solutions is not immediately straightforward because ionic conductivity introduces additional spectral features [44,50,85] and complicates determination of the static dielectric constant [85,86]. For example, we expect that ion-free aqueous systems such as dilute sugar solutions could be more suitable [87]. In general, based on the trends observed here for pure water, a strengthened (weakened) hydrogen-bond network should be described by slower (faster) Debye reorientation dynamics [88] and reduced (increased) Drude-Smith spectral weight.

Finally, by combining experiments described by a Debye plus Drude-Smith model with *ab-initio*-based simulations, a plausible and complementary picture of the low-frequency terahertz response of liquid water emerged. On the theory side, charge transfer contributes to polarization [13] and, in turn, polarization influences the dielectric response of water in the ~0.1-1 THz range, as shown in Figure 4b. In the same spectral range, the Drude-Smith model could describe those water molecules with an instantaneous imbalance between donated and accepted hydrogen bonds, which are characterised by an asymmetric charge transfer. Therefore, we suggest that the low-frequency terahertz spectrum of liquid water might stem not only from those molecular reorientations captured by the Debye model, but also from charge-transfer related local polarization effects that could be described by an empirical Drude-Smith approach.

## SUPPLEMENTARY MATERIAL

See the supplementary material for the calculated hydrogen-bond population tables, additional simulation-protocol details, and convergence tests of the dipole autocorrelation function.

## ACKNOWLEDGMENTS

FN acknowledges financial support by the DFG (509442914), University of Southampton (Startup package), and Ruhr University Bochum (Novizenprämie, RDSS). AAH acknowledges the funding received by the European Research Council (ERC) under the European Union's Horizon 2020 research and innovation program (Grant No. 101043272 - HyBOP). The views and opinions expressed are those of the authors only and do not necessarily reflect those of the European Union or the European Research Council Executive Agency. Neither the European Union nor the granting authority can be held responsible for them. GC is thankful to CINECA for awards under the ISCRA initiative, for the availability of high-performance computing resources and support.

## AUTHOR DECLARATIONS

### Conflict of Interest

The authors have no conflicts to disclose.

### Author Contributions

Lilian Najm Alsayed: Investigation; Formal analysis; Validation; Visualization; Writing – review & editing. Florian Pabst: Investigation; Methodology; Formal analysis; Visualization; Writing – review & editing. Giuseppe Cassone: Investigation; Methodology; Formal analysis; Writing – review & editing. Ali Hassanali: Conceptualization; Methodology; Project administration; Supervision; Writing – review & editing. Fabio Novelli: Investigation; Conceptualization; Methodology; Formal analysis; Visualization; Project administration; Supervision; Writing – original draft; Writing – review & editing.

## DATA AVAILABILITY

The data that support the findings of this study are available from the authors upon reasonable request.

## REFERENCES

[1] E. Arunan, G.R. Desiraju, R.A. Klein, J. Sadlej, S. Scheiner, I. Alkorta, D.C. Clary, R.H. Crabtree, J.J. Dannenberg, P. Hobza, H.G. Kjaergaard, A.C. Legon, B. Mennucci, and D.J.

Nesbitt, "Definition of the hydrogen bond (IUPAC Recommendations 2011)," Pure and Applied Chemistry **83**(8), 1637–1641 (2011).

[2] A. Chimarro-Contreras, Y. Lopez-Revelo, J. Cardenas-Gamboa, and T. Terencio, "Insights into the Effect of Charges on Hydrogen Bonds," Int. J. Mol. Sci. **25**(3), 1613 (2024).

[3] A.J. Lee, and S.W. Rick, "The effects of charge transfer on the properties of liquid water," J. Chem. Phys. **134**(18), 184507 (2011).

[4] D. Ojha, K. Karhan, and T.D. Kühne, "On the Hydrogen Bond Strength and Vibrational Spectroscopy of Liquid Water," Sci. Rep. **8**(1), 16888 (2018).

[5] D. Ben-Amotz, "Electric buzz in a glass of pure water," Science **376**(6595), 800–801 (2022).

[6] C. Schran, O. Marsalek, and T.E. Markland, "Unravelling the influence of quantum proton delocalization on electronic charge transfer through the hydrogen bond," Chem. Phys. Lett. **678**, 289–295 (2017).

[7] R. Kumar, J.R. Schmidt, and J.L. Skinner, "Hydrogen bonding definitions and dynamics in liquid water," J. Chem. Phys. **126**(20), 204107 (2007).

[8] H. Ito, T. Hasegawa, and Y. Tanimura, "Effects of Intermolecular Charge Transfer in Liquid Water on Raman Spectra," Journal of Physical Chemistry Letters **7**(20), 4147–4151 (2016).

[9] H. Torii, "Intermolecular electron density modulations in water and their effects on the far-infrared spectral profiles at 6 THz," Journal of Physical Chemistry B **115**(20), 6636–6643 (2011).

[10] H. Torii, "Cooperative contributions of the intermolecular charge fluxes and intramolecular polarizations in the far-infrared spectral intensities of liquid water," J. Chem. Theory Comput. **10**(3), 1219–1227 (2014).

[11] M. Flór, D.M. Wilkins, M. de la Puente, D. Laage, G. Cassone, A. Hassanali, and S. Roke, "Dissecting the hydrogen bond network of water: Charge transfer and nuclear quantum effects," Science **386**(6726), eads4369 (2024).

[12] A. Amadeo, M.F. Torre, K. Mráziková, F. Saija, S. Trusso, J. Xie, M. Tommasini, and G. Cassone, "Hydrogen Bonds under Electric Fields with Quantum Accuracy," J. Phys. Chem. A **129**(18), 4077–4092 (2025).

[13] X. Ge, and D. Lu, "Molecular polarizability of water from local dielectric response theory," Phys. Rev. B **96**(7), 075114 (2017).

[14] S. Pullanchery, S. Kulik, B. Rehl, A. Hassanali, and S. Roke, "Charge transfer across C–H···O hydrogen bonds stabilizes oil droplets in water," Science **374**(6573), 1366–1370 (2021).

[15] E. Poli, K.H. Jong, and A. Hassanali, "Charge transfer as a ubiquitous mechanism in determining the negative charge at hydrophobic interfaces," Nat. Commun. **11**(1), 901 (2020).

[16] P. Gasparotto, A.A. Hassanali, and M. Ceriotti, "Probing Defects and Correlations in the Hydrogen-Bond Network of ab Initio Water," J. Chem. Theory Comput. **12**(4), 1953–1964 (2016).

[17] Z. Wang, Z. Chen, J. Fang, S. Li, W. Zhou, and H. Qiu, "Charge transfer at air-water interfaces: A machine learning potential-based molecular dynamics study," Journal of Chemical Physics **162**(24), 244705 (2025).

[18] G. Amante, F. Panzera, G. Centi, J. Xie, A. Hassanali, A.M. Saitta, and G. Cassone, "Interfacial Electric Fields in Water Nanodroplets are Weakly Dependent on Curvature and pH," arXiv:2604.14784 (2026).

[19] J. Takeda, K. Yoshioka, Y. Minami, and I. Katayama, "Nanoscale electron manipulation in metals with intense THz electric fields," J. Phys. D Appl. Phys. **51**(10), 103001 (2018).

[20] M. Walther, D.G. Cooke, C. Sherstan, M. Hajar, M.R. Freeman, and F.A. Hegmann, "Terahertz conductivity of thin gold films at the metal-insulator percolation transition," Phys. Rev. B **76**(12), 125408 (2007).

[21] K. Shimakawa, and S. Kasap, "Dynamics of Carrier Transport in Nanoscale Materials: Origin of Non-Drude Behavior in the Terahertz Frequency Range," Applied Sciences **6**(2), 50 (2016).

[22] T.L. Cocker, L. V. Titova, S. Fourmaux, H.-C. Bandulet, D. Brassard, J.-C. Kieffer, M.A. El Khakani, and F.A. Hegmann, "Terahertz conductivity of the metal-insulator transition in a nanogranular VO2 film," Appl. Phys. Lett. **97**(22), 221905 (2010).

[23] G. Li, D. Li, Z. Jin, and G. Ma, "Photocarriers dynamics in silicon wafer studied with optical-pump terahertz-probe spectroscopy," Opt. Commun. **285**(20), 4102–4106 (2012).

[24] D. Tsokkou, A. Othonos, and M. Zervos, "Carrier dynamics and conductivity of SnO2 nanowires investigated by time-resolved terahertz spectroscopy," Appl. Phys. Lett. **100**(13), 133101 (2012).

[25] H. Ahn, Y.-P. Ku, Y.-C. Wang, C.-H. Chuang, S. Gwo, and C.-L. Pan, "Terahertz spectroscopic study of vertically aligned InN nanorods," Appl. Phys. Lett. **91**(16), 163105 (2007).

[26] R. Lovrinčić, and A. Pucci, "Infrared optical properties of chromium nanoscale films with a phase transition," Phys. Rev. B **80**(20), 205404 (2009).

[27] D.G. Cooke, A.N. MacDonald, A. Hryciw, J. Wang, Q. Li, A. Meldrum, and F.A. Hegmann, "Transient terahertz conductivity in photoexcited silicon nanocrystal films," Phys. Rev. B **73**(19), 193311 (2006).

[28] J. Kim, I. Maeng, J. Jung, H. Song, J.-H. Son, K. Kim, J. Lee, C.-H. Kim, G. Chae, M. Jun, Y. Hwang, S. Jeong Lee, J.-M. Myoung, and H. Choi, "Terahertz time-domain measurement of non-Drude conductivity in silver nanowire thin films for transparent electrode applications," Appl. Phys. Lett. **102**(1), 011109 (2013).

[29] N. Smith, "Classical generalization of the Drude formula for the optical conductivity," Phys. Rev. B **64**(15), 155106 (2001).

[30] T.L. Cocker, D. Baillie, M. Buruma, L. V. Titova, R.D. Sydora, F. Marsiglio, and F.A. Hegmann, "Microscopic origin of the Drude-Smith model," Phys. Rev. B **96**(20), 205439 (2017).

[31] P. Kužel, and H. Němec, "Terahertz Spectroscopy of Nanomaterials: a Close Look at Charge-Carrier Transport," Adv. Opt. Mater. **8**(3), 1900623 (2020).

[32] W.C. Chen, and R.A. Marcus, "The Drude-Smith Equation and Related Equations for the Frequency-Dependent Electrical Conductivity of Materials: Insight from a Memory Function Formalism," ChemPhysChem **22**(16), 1667–1674 (2021).

[33] C. Rønne, and S.R. Keiding, "Low frequency spectroscopy of liquid water using THz-time domain spectroscopy," J. Mol. Liq. **101**(1–3), 199–218 (2002).

[34] C. Rønne, P.-O. Åstrand, and S.R. Keiding, "THz Spectroscopy of Liquid H2O and D2O," Phys. Rev. Lett. **82**(14), 2888–2891 (1999).
[35] H.J. Liebe, G.A. Hufford, and T. Manabe, "A model for the complex permittivity of water at frequencies below 1 THz," Int. J. Infrared Millimeter Waves **12**(7), 659–675 (1991).
[36] H. Yada, M. Nagai, and K. Tanaka, "Origin of the fast relaxation component of water and heavy water revealed by terahertz time-domain attenuated total reflection spectroscopy," Chem. Phys. Lett. **464**(4–6), 166–170 (2008).
[37] B. Kutus, A. Shalit, P. Hamm, and J. Hunger, "Dielectric response of light, heavy and heavy-oxygen water: isotope effects on the hydrogen-bonding network's collective relaxation dynamics," Physical Chemistry Chemical Physics **23**(9), 5467–5473 (2021).
[38] F. Alvarez, A. Arbe, and J. Colmenero, "The Debye's model for the dielectric relaxation of liquid water and the role of cross-dipolar correlations. A MD-simulations study," J. Chem. Phys. **159**(13), 134505 (2023).
[39] R. Buchner, J. Barthel, and J. Stauber, "The dielectric relaxation of water between 0°C and 35°C," Chem. Phys. Lett. **306**(1–2), 57–63 (1999).
[40] W.J. Ellison, "Permittivity of Pure Water, at Standard Atmospheric Pressure, over the Frequency Range 0–25 THz and the Temperature Range 0–100 °C," J. Phys. Chem. Ref. Data **36**(1), 1–18 (2007).
[41] K. Shiraga, K. Tanaka, T. Arikawa, S. Saito, and Y. Ogawa, "Reconsideration of the relaxational and vibrational line shapes of liquid water based on ultrabroadband dielectric spectroscopy," Physical Chemistry Chemical Physics **20**(41), 26200–26209 (2018).
[42] D.C. Elton, "The origin of the Debye relaxation in liquid water and fitting the high frequency excess response," Physical Chemistry Chemical Physics **19**(28), 18739–18749 (2017).
[43] J. Hunger, L. Liu, K.-J. Tielrooij, M. Bonn, and H. Bakker, "Vibrational and orientational dynamics of water in aqueous hydroxide solutions," J. Chem. Phys. **135**(12), 124517 (2011).
[44] V. Balos, S. Imoto, R.R. Netz, M. Bonn, D.J. Bonthuis, Y. Nagata, and J. Hunger, "Macroscopic conductivity of aqueous electrolyte solutions scales with ultrafast microscopic ion motions," Nat. Commun. **11**(1), 1611 (2020).
[45] A. Taschin, P. Bartolini, L. Caminiti, A. Boschetti, S. Fanetti, and R. Torre, "Broadband THz time-domain spectroscopy of water: Enhanced cooperativity of hydrogen-bond dynamics in the Debye wing," J. Mol. Liq. **460**, 129935 (2026).
[46] P. Lunkenheimer, S. Emmert, R. Gulich, M. Köhler, M. Wolf, M. Schwab, and A. Loidl, "Electromagnetic-radiation absorption by water," Phys. Rev. E **96**(6), 062607 (2017).
[47] W. Schirmacher, G. Diezemann, and C. Ganter, "Harmonic Vibrational Excitations in Disordered Solids and the 'Boson Peak,'" Phys. Rev. Lett. **81**(1), 136–139 (1998).
[48] T. Fukasawa, T. Sato, J. Watanabe, Y. Hama, W. Kunz, and R. Buchner, "Relation between Dielectric and Low-Frequency Raman Spectra of Hydrogen-Bond Liquids," Phys. Rev. Lett. **95**(19), 197802 (2005).
[49] A.A. Volkov, and S. V. Chuchupal, "THz insight into Debye relaxation in liquid water: Link to self-diffusion and viscosity," J. Chem. Phys. **163**(22), 224502 (2025).
[50] N.Q. Vinh, M.S. Sherwin, S.J. Allen, D.K. George, A.J. Rahmani, and K.W. Plaxco, "High-precision gigahertz-to-terahertz spectroscopy of aqueous salt solutions as a probe of the

femtosecond-to-picosecond dynamics of liquid water," J. Chem. Phys. **142**(16), 164502 (2015).
[51] N. Penkov, N. Shvirst, V. Yashin, and E. Fesenko, "Terahertz Spectroscopy Applied for Investigation of Water Structure," Journal of Physical Chemistry B **119**(39), 12664–12670 (2015).
[52] E.M. Adams, H. Hao, I. Leven, M. Rüttermann, H. Wirtz, M. Havenith, and T. Head-Gordon, "Proton Traffic Jam: Effect of Nanoconfinement and Acid Concentration on Proton Hopping Mechanism," Angewandte Chemie **133**(48), 25623–25631 (2021).
[53] F. Sebastiani, S.L.P. Wolf, B. Born, T.Q. Luong, H. Cölfen, D. Gebauer, and M. Havenith, "Water Dynamics from THz Spectroscopy Reveal the Locus of a Liquid–Liquid Binodal Limit in Aqueous $CaCO_3$ Solutions," Angew. Chem. Int. Ed. **56**(2), 490–495 (2017).
[54] C. Hölzl, H. Forbert, and D. Marx, "Dielectric relaxation of water: Assessing the impact of localized modes, translational diffusion, and collective dynamics," Physical Chemistry Chemical Physics **23**(37), 20875–20882 (2021).
[55] F. Pabst, and S. Baroni, "How Salt Solvation Slows Water Dynamics While Blue-Shifting Its Dielectric Spectrum," J. Phys. Chem. Lett. **16**(31), 7915–7920 (2025).
[56] A. Beneduci, "Which is the effective time scale of the fast Debye relaxation process in water?," J. Mol. Liq. **138**(1–3), 55–60 (2008).
[57] M. Heyden, J. Sun, S. Funkner, G. Mathias, H. Forbert, M. Havenith, and D. Marx, "Dissecting the THz spectrum of liquid water from first principles via correlations in time and space," Proc. Natl. Acad. Sci. U. S. A. **107**(27), 12068–12073 (2010).
[58] D. Sharma, B. Das, and A. Chandra, "Terahertz Spectrum of Water at Varying Temperatures from 260 to 340 K: Contributions from Permanent and Induced Dipole Correlations at Different Length Scales," Journal of Physical Chemistry B **127**(30), 6714–6725 (2023).
[59] F. Novelli, "Terahertz spectroscopy of thick and diluted water solutions," Opt. Express **32**(7), 11041 (2024).
[60] F. Novelli, L. Spina, A. Rubano, L. Cattaneo, D. Paparo, F. Ciuchi, and M.P. De Santo, "Intense TeraHertz Time Domain Spectroscopy of Water-based Liquid Crystals," IEEE Trans. Terahertz Sci. Technol. (10.1109/TTHZ.2026.3690056), 1–10 (2026).
[61] F. Novelli, "Terahertz Transmission through a Gold Mirror or Electrode," Materials **17**(16), 3942 (2024).
[62] C.G. Malmberg, "Dielectric constant of deuterium oxide," J. Res. Natl. Bur. Stand. **60**(6), 609 (1958).
[63] D.G. Archer, and P. Wang, "The Dielectric Constant of Water and Debye-Hückel Limiting Law Slopes," J. Phys. Chem. Ref. Data **19**(2), 371–411 (1990).
[64] G.A. Vidulich, D.F. Evans, and R.L. Kay, "The dielectric constant of water and heavy water between 0 and 40°," J. Phys. Chem. **71**(3), 656–662 (1967).
[65] J. Lloyd-Hughes, and T.I. Jeon, "A review of the terahertz conductivity of bulk and nano-materials," J. Infrared Millim. Terahertz Waves **33**(9), 871–925 (2012).
[66] A.P. Thompson, H.M. Aktulga, R. Berger, D.S. Bolintineanu, W.M. Brown, P.S. Crozier, P.J. in 't Veld, A. Kohlmeyer, S.G. Moore, T.D. Nguyen, R. Shan, M.J. Stevens, J. Tranchida, C. Trott, and S.J. Plimpton, "LAMMPS - a flexible simulation tool for particle-based

materials modeling at the atomic, meso, and continuum scales," Comput. Phys. Commun. **271**, 108171 (2022).

[67] Y. Zhao, and D.G. Truhlar, "The M06 suite of density functionals for main group thermochemistry, thermochemical kinetics, noncovalent interactions, excited states, and transition elements: two new functionals and systematic testing of four M06-class functionals and 12 other functionals," Theor. Chem. Acc. **120**(1–3), 215–241 (2008).

[68] P. Zhang, and X. Xu, "Propensity of Water Self-Ions at Air(Oil)–Water Interfaces Revealed by Deep Potential Molecular Dynamics with Enhanced Sampling," Langmuir **41**(5), 3675–3683 (2025).

[69] P. Zhang, "Oil-Water Interface Study Resources," Https://Github.Com/Zhang-Pchao/OilWaterInterface, accessed 15 July 2026 (n.d.).

[70] Y. Li, A. Gomez, K. Cai, C. Zhang, L. Fu, W. Jia, Y.M.Y. Feldman, O. Blumer, J. Higer, B. Hirshberg, S. Xu, A. Kohlmeyer, and R. Car, "fix pimd/langevin: An efficient implementation of path integral molecular dynamics in LAMMPS," J. Chem. Phys. **164**(14), 144119 (2026).

[71] S.L. Bore, and F. Paesani, "Realistic phase diagram of water from 'first principles' data-driven quantum simulations," Nat. Commun. **14**(1), 3349 (2023).

[72] J. Zeng, D. Zhang, D. Lu, P. Mo, Z. Li, Y. Chen, M. Rynik, L. Huang, Z. Li, S. Shi, Y. Wang, H. Ye, P. Tuo, J. Yang, Y. Ding, Y. Li, D. Tisi, Q. Zeng, H. Bao, Y. Xia, J. Huang, K. Muraoka, Y. Wang, J. Chang, F. Yuan, S.L. Bore, C. Cai, Y. Lin, B. Wang, J. Xu, J.X. Zhu, C. Luo, Y. Zhang, R.E.A. Goodall, W. Liang, A.K. Singh, S. Yao, J. Zhang, R. Wentzcovitch, J. Han, J. Liu, W. Jia, D.M. York, E. Weinan, R. Car, L. Zhang, and H. Wang, "DeePMD-kit v2: A software package for deep potential models," Journal of Chemical Physics **159**(5), 54801 (2023).

[73] C. Malosso, N. Manko, M.G. Izzo, S. Baroni, and A. Hassanali, "Evidence of ferroelectric features in low-density supercooled water from ab initio deep neural-network simulations," Proceedings of the National Academy of Sciences **121**(32), e2407295121 (2024).

[74] J. Sun, A. Ruzsinszky, and J.P. Perdew, "Strongly Constrained and Appropriately Normed Semilocal Density Functional," Phys. Rev. Lett. **115**(3), 036402 (2015).

[75] A. Luzar, and D. Chandler, "Effect of Environment on Hydrogen Bond Dynamics in Liquid Water," Phys. Rev. Lett. **76**(6), 928–931 (1996).

[76] N. Marzari, and D. Vanderbilt, "Maximally localized generalized Wannier functions for composite energy bands," Phys. Rev. B **56**(20), 12847–12865 (1997).

[77] R.Z. Khaliullin, A.T. Bell, and M. Head-Gordon, "Electron Donation in the Water–Water Hydrogen Bond," Chemistry – A European Journal **15**(4), 851–855 (2009).

[78] T.D. Kühne, and R.Z. Khaliullin, "Electronic signature of the instantaneous asymmetry in the first coordination shell of liquid water," Nat. Commun. **4**(1), 1450 (2013).

[79] R. Vácha, S.W. Rick, P. Jungwirth, A.G.F. de Beer, H.B. de Aguiar, J.-S. Samson, and S. Roke, "The Orientation and Charge of Water at the Hydrophobic Oil Droplet–Water Interface," J. Am. Chem. Soc. **133**(26), 10204–10210 (2011).

[80] H.R. Zelsmann, "Temperature dependence of the optical constants for liquid H2O and D2O in the far IR region," J. Mol. Struct. **350**(2), 95–114 (1995).

[81] I. Popov, P. Ben Ishai, A. Khamzin, and Y. Feldman, "The mechanism of the dielectric relaxation in water," Physical Chemistry Chemical Physics **18**(20), 13941–13953 (2016).

[82] V.G. Artemov, "A unified mechanism for ice and water electrical conductivity from direct current to terahertz," Physical Chemistry Chemical Physics **21**(15), 8067–8072 (2019).
[83] A.A. Volkov, V.G. Artemov, and A. V. Pronin, "A radically new suggestion about the electrodynamics of water: Can the pH index and the Debye relaxation be of a common origin?," EPL (Europhysics Letters) **106**(4), 46004 (2014).
[84] S. Liu, and J. Yuan, "Unveiling dielectric relaxation and shear rheology in water," Phys. Rev. Res. **8**(2), 023266 (2026).
[85] T. Chen, G. Hefter, and R. Buchner, "Dielectric Spectroscopy of Aqueous Solutions of KCl and CsCl," J. Phys. Chem. A **107**(20), 4025–4031 (2003).
[86] K.J. Tielrooij, R.L.A. Timmer, H.J. Bakker, and M. Bonn, "Structure Dynamics of the Proton in Liquid Water Probed with Terahertz Time-Domain Spectroscopy," Phys. Rev. Lett. **102**(19), 198303 (2009).
[87] M.J. Zapanta, F. Novelli, A. Postelmans, and W. Saeys, "High-intensity terahertz time-domain spectroscopy (HI-THz-TDS) reveals hydration behavior of lactose in the dilute regime," Spectrochim. Acta A Mol. Biomol. Spectrosc. **362**, 128171 (2026).
[88] F. Novelli, and E. Adams, "Terahertz Spectroscopy of Water and Proteins: Methodology and Insights," in *Dynamics of Biomolecules Studied by Dielectric Spectroscopy*, edited by S. Cerveny and J. Swenson, (Springer Cham, 2026), pp. 189–214.